\documentclass{article}

 \usepackage[preprint]{neurips_2026}

\usepackage{amsmath}
\usepackage{graphicx}
\usepackage[utf8]{inputenc} 
\usepackage[T1]{fontenc}    
\usepackage{hyperref}       
\usepackage{xurl}
\usepackage[T1]{fontenc}
\let\oldref\ref
\renewcommand{\ref}[1]{\mbox{\oldref{#1}}}
\let\oldcitep\citep
\renewcommand{\citep}[1]{\mbox{\oldcitep{#1}}}
\let\oldcite\cite
\renewcommand{\cite}[1]{\mbox{\oldcite{#1}}}
\usepackage{url}            
\usepackage{booktabs}       
\usepackage{amsfonts}       
\usepackage{nicefrac}       
\usepackage{microtype}      
\usepackage{xcolor}         
\usepackage{algorithm}      
\usepackage{algpseudocode}  
\usepackage{placeins}       
\usepackage{tikz}
\usepackage{fontawesome5}
\usepackage{standalone}
\usetikzlibrary{
  arrows.meta,
  calc,
  positioning,
  shapes.geometric,
  shapes.misc,
  fit,
  backgrounds,
  shadows
}
\usepackage{soul}
\usepackage{multirow}
\definecolor{srcblue}{HTML}{3B82F6}        
\definecolor{tgtgreen}{HTML}{10B981}       
\definecolor{rednode}{HTML}{EF4444}        
\definecolor{protoviolet}{HTML}{8B5CF6}    
\definecolor{alignorange}{HTML}{F59E0B}    
\definecolor{modelgray}{HTML}{64748B}      
\definecolor{bgcard}{HTML}{F8FAFC}         
\definecolor{bordergray}{HTML}{CBD5E1}     

\title{Prototype-Guided Latent Alignment for Data-Efficient Fine-Tuning of Molecular Foundation Models }

\author{
    Rushikesh Pawar \qquad
    Harshit Rawat \qquad
    Ayush Kumar \qquad
    Phani Motamarri$^{*}$ \\ \\
    Department of Computational and Data Sciences\\ Indian Institute of Science, Bangalore, 560012 \\ \texttt{$^{*}$phanim@iisc.ac.in}    
}

\begin{document}

\maketitle

\begin{abstract}
    Machine learning interatomic potentials (MLIPs) have transformed materials discovery by leveraging Graph Neural Networks (GNN) to predict material properties with near density functional theory (DFT) accuracy. While large-scale pretrained foundation models offer transferable baseline predictions, they frequently struggle to generalise to out-of-distribution (OOD) target systems, a common challenge encountered when modelling complex or chemically diverse material systems. Fine-tuning on target-domain data is the standard remedy, but the high computational cost of generating DFT-labelled configurations confines this adaptation to data-scarce regimes. In such settings, the over-parameterised nature of modern GNNs amplifies overfitting, severely degrading predictive performance on the target domain. To address this challenge, we propose a prototype-based alignment approach for data-efficient fine-tuning of MLIPs. Our approach systematically identifies local structural similarities between the source and target domains by grouping atoms with similar local chemical environments through their latent representations. The energy contribution of each target-domain atom is then aligned to that of its corresponding source-domain prototype, introducing an inductive bias that anchors the fine-tuned representations to the pretrained source structure. This alignment encourages effective reuse of learned interatomic interactions and substantially improves generalization under data-scarce conditions, without imposing restrictive assumptions on the target chemistry. We rigorously evaluate our method across multiple molecular target systems in the rMD17 benchmark using two widely adopted GNN architectures, equivariant MACE and invariant SchNet architectures, across varying data budgets. We further extend evaluation to large-scale universal models by applying prototype-guided fine-tuning to the MACE-OFF series of foundation models on the SPICE dataset. Across all settings, our approach consistently improves predictive accuracy in the low-data regime, reducing energy mean absolute error (MAE) by up to 18\% relative to standard fine-tuning baselines.
\end{abstract}

\section{Introduction}
Machine learning interatomic potentials (MLIPs) have fundamentally transformed atomistic modelling by bridging the long-standing trade-off between accuracy and computational efficiency. Traditional quantum mechanical approaches, such as density functional theory (DFT), provide highly accurate predictions of atomic energies and forces, but are computationally expensive for large-scale simulations. In contrast, classical force fields offer high efficiency at the cost of limited accuracy and transferability. MLIPs overcome this limitation by learning high-dimensional potential energy surfaces directly from DFT data, achieving near-DFT accuracy while reducing the computational cost \citep{WANG2024109673}. Early developments in MLIPs relied on handcrafted descriptors \citep{behler2011atom, Bart_k_2013} combined with regression models, including high-dimensional neural network potentials (HDNNP) \citep{Behler_2014}, Gaussian Approximation Potentials (GAP) \citep{PhysRevLett.104.136403} , Moment Tensor potentials (MTP) \citep{doi:10.1137/15M1054183}, and Spectral Neighbour Analysis Potentials (SNAP) \citep{THOMPSON2015316}. These approaches encode local atomic environments into invariant feature vectors before learning energy mappings. While effective, their performance depends heavily on descriptor design, limiting flexibility and generalization. Recent advances have shifted towards  graph neural networks (GNNs) architectures, such as SchNet \citep{10.1063/1.5019779}, CGCNN \citep{Xie2017CrystalGC}, PaiNN \citep{schutt2021equivariant}, NequIP \citep{batzner20223}, M3GNET \citep{Chen2022AUG}, MACE \citep{batatia2022mace} and GRACE \citep{Lysogorskiy2025GraphAC} which learn representations directly from atomic coordinates while respecting physical symmetries. These models differ in their treatment of geometry, ranging from invariant to equivariant formulations. SchNet represents an early class of invariant message passing networks where interactions depend only on interatomic distances, ensuring rotational invariance but limiting the ability to model directional dependencies. While recent models like MACE employ higher-order equivariant message passing enabling features to transform consistently under rotations and thereby capturing richer geometric structure.


Building on these advances, recent efforts have shifted toward foundation models (FM) trained on large-scale, diverse datasets spanning broad chemical and configurational spaces. Notable examples include the Materials Project~\citep{10.1063/1.4812323}, which provides extensive density functional theory (DFT) calculations for crystalline materials; OMat~\citep{BarrosoLuque2024OpenM2}, a large and diverse collection of atomistic simulations designed to support foundation models across materials chemistry; and SPICE~\citep{Eastman2022SPICEAD}, which focuses on high-quality quantum chemical data for small molecules, including energies, forces, and conformational variations. Together, these datasets enable the training of models with improved transferability across both molecular and materials system. Models such as MACE, CHGNet \citep{deng2023chgnet}, M3GNet \citep{chen2022universal} and related architectures aim to provide transferable potentials that generalize across systems without retraining. Prominent examples of such large-scale foundation models include MACE-MP-0 \citep{Batatia2023AFM}, MACE-OFF \citep{Kovacs2023MACEOFFST} trained on the Materials Project and SPICE datasets respectively. However, empirical evaluations reveal that their performance can degrade significantly when applied to out-of-distribution systems \citep{Focassio_2024}. Consequently, fine-tuning foundation models on target system has become the standard paradigm for adapting MLIPs to new systems \citep{Radova2025FinetuningFM}.  Despite their strong transferability, adapting these models is fundamentally constrained by the high cost of generating labeled data via DFT calculations. As a result, maximizing sample efficiency during fine-tuning is not merely desirable, but critical for practical applicability. Prior work on leveraging foundationMLIPs has largely focused on improving computational efficiency at deployment time, rather than addressing data efficiency during fine-tuning . One line of work distills large pretrained models into smaller, more efficient surrogates \citep{Gardner2025DistillationOA}, relying on additional synthetic data generated by perturbing the training data. Another approach \citep{Radova2025FinetuningFM} employs kernel-based methods where the optimization problem has a closed form solution, thereby eliminating the need for expensive gradient-based optimization.

While prior approaches primarily improve computational efficiency, they do not address the central bottleneck of limited labeled data in the target system of interest. We instead focus on improving the sample efficiency of fine-tuning by leveraging a fundamental property of MLIPs: the locality of atomic interactions, which implies that similar local environments across different systems should exhibit similar energies \citep{PhysRevLett.98.146401}. Despite being implicit in existing models, this cross-system structural similarity is not explicitly exploited during fine-tuning. To address this gap, we propose a prototype guided latent alignment framework that leverages structural representations learned by a pretrained model to guide learning on a target system. Specifically, we construct prototypes representing local atomic environments in the latent space, along with associated energy statistics, and introduce a Nodal Energy Alignment Loss (NEAL) that enforces consistency between predicted atomic energies and their matched prototypes during fine-tuning. This alignment acts as an inductive bias that stabilizes learning in low-data regimes while allowing flexible learning in later stages. We evaluate our method across both proxy and large-scale pretrained MLIPs, including SchNet and MACE, on diverse molecular systems and data budgets, demonstrating consistent improvements over standard fine-tuning, with the largest gains in low-data settings. MACE and SchNet are strategically chosen as they are two prominent architectures representing the invariant and equivariant ends of the molecular GNN spectrum, respectively. By demonstrating the performance of our framework on both architectures, we demonstrate the generalization of our method across a broad class of graph neural network models. Our main contributions are: (1) identifying cross-system local structural similarity as a key inductive bias for data-efficient fine-tuning of MLIPs, (2) introducing a prototype-guided latent alignment mechanism with NEAL to explicitly enforce this bias during fine-tuning, and (3) demonstrating robust and consistent empirical gains across architectures and datasets under severe data constraints.

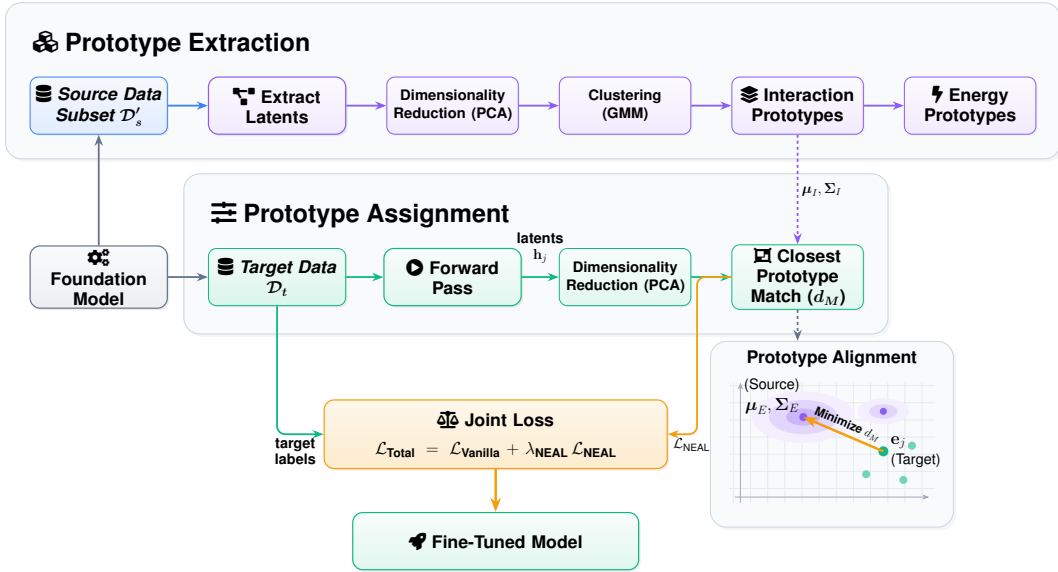
\begin{figure}[htbp]
    \centering
    \resizebox{\textwidth}{!}{\begin{tikzpicture}[
    >=Stealth,
    text=black,
    node distance=1.6cm and 1.4cm, 
    box/.style={
        draw=#1, top color=#1!3, bottom color=#1!12, rounded corners=10pt,
        minimum width=4.8cm, minimum height=2.0cm, text width=4.4cm, 
        align=center, line width=1.2pt,
        font=\sffamily\LARGE\bfseries, 
        drop shadow={opacity=0.15, shadow xshift=2pt, shadow yshift=-2pt}
    },
    smallBox/.style={
        draw=#1, top color=#1!3, bottom color=#1!12, rounded corners=8pt,
        minimum width=4.6cm, minimum height=2.0cm, text width=4.2cm, 
        align=center, line width=1pt,
        font=\sffamily\LARGE, 
        drop shadow={opacity=0.1, shadow xshift=1.5pt, shadow yshift=-1.5pt}
    },
    dataBox/.style={
        draw=#1, top color=#1!5, bottom color=#1!15, rounded corners=8pt,
        minimum width=4.8cm, minimum height=2.0cm, text width=4.4cm, 
        align=center, line width=1pt,
        font=\sffamily\LARGE\itshape, 
        drop shadow={opacity=0.15, shadow xshift=1.5pt, shadow yshift=-1.5pt}
    },
    lossBox/.style={
        draw=#1, top color=#1!8, bottom color=#1!20, rounded corners=10pt,
        minimum width=12.0cm, minimum height=2.4cm, text width=11.5cm, 
        align=center, line width=1.2pt,
        font=\sffamily\LARGE\bfseries, 
        drop shadow={opacity=0.15, shadow xshift=2pt, shadow yshift=-2pt}
    },
    cardBg/.style={
        fill=bgcard, draw=bordergray, rounded corners=16pt,
        line width=1pt, inner sep=24pt, 
        drop shadow={opacity=0.08, shadow xshift=3pt, shadow yshift=-3pt}
    },
    sectionLabel/.style={
        font=\sffamily\Huge\bfseries, text=black, align=left 
    },
    arr/.style={->, line width=1.8pt, color=#1, rounded corners=8pt},
    darr/.style={->, line width=1.8pt, color=#1, dashed, rounded corners=8pt},
    note/.style={font=\sffamily\Large, text=black, align=center} 
]

\node[box=modelgray] (UM) at (0, 2.0) {\faCogs\\Foundation Model};

\node[dataBox=srcblue] (srcData) at (0, 8.0) {\faDatabase\ \textbf{Source Data Subset} $\boldsymbol{\mathcal{D}'_s}$};

\node[box=protoviolet, right=of srcData] (extractLatent) {\faProjectDiagram\ Extract\\Latents};
\node[smallBox=protoviolet, right=of extractLatent] (pca) {{\Large\textbf{Dimensionality\\[-2pt]Reduction (PCA)}}};
\node[smallBox=protoviolet, right=of pca] (gmm) {{\Large\textbf{Clustering\\[-2pt](GMM)}}};
\node[smallBox=protoviolet, right=of gmm] (intProto) {\faLayerGroup\ \textbf{Interaction\\Prototypes}};
\node[smallBox=protoviolet, right=of intProto] (engProto) {\faBolt\ \textbf{Energy\\Prototypes}};

\node[sectionLabel, above=0.6cm of srcData.north west, anchor=south west] 
    (protoTitle) {\faIcon{cubes}\ Prototype Extraction};
\begin{scope}[on background layer]
    \node[cardBg, fit=(protoTitle)(srcData)(engProto)] (protoCard) {};
\end{scope}

\node[dataBox=tgtgreen, right=of UM] (tgtData) {\faDatabase\ \textbf{Target Data}\\$\boldsymbol{\mathcal{D}_t}$};

\node[box=tgtgreen] (fwdTarget) at (pca |- tgtData) {\faPlayCircle\ Forward\\Pass};
\node[smallBox=tgtgreen] (tgtPca) at (gmm |- tgtData) {{\Large\textbf{Dimensionality\\[-2pt]Reduction (PCA)}}};
\node[smallBox=tgtgreen] (matchProto) at (intProto |- tgtData) {\faObjectGroup\ \textbf{Closest Prototype Match ($\boldsymbol{d_M}$)}};

\node[sectionLabel, above=0.6cm of tgtData.north west, anchor=south west] 
    (alignTitle) {\faIcon{sliders-h}\ Prototype Assignment};
\begin{scope}[on background layer]
    \node[cardBg, fit=(alignTitle)(tgtData)(matchProto)] (alignCard) {};
\end{scope}

\node[lossBox=alignorange] (combinedLoss) at ($(fwdTarget)!0.25!(tgtPca) - (0, 5.5cm)$) 
    {\faBalanceScale\ Joint Loss\\[8pt]
    $\mathcal{L}_{\text{Total}} = \mathcal{L}_{\text{Vanilla}}  + \mathcal{\lambda_{\text{NEAL}}}\,\mathcal{L}_{\text{NEAL}}$};

\node[box=tgtgreen, minimum width=10.0cm, text width=9.5cm, below=1.5cm of combinedLoss] (output) 
    {\faRocket\ Fine-Tuned Model}; 

\node[cardBg, minimum width=8.5cm, minimum height=6.5cm] (schematic) at ([xshift=1.2cm]matchProto |- combinedLoss) {};

\begin{scope}
    \node[font=\sffamily\fontsize{24}{28}\selectfont\bfseries, text=black, anchor=north, yshift=-6pt] at (schematic.north) {\LARGE Prototype Alignment}; 

    \draw[step=0.6cm, modelgray!15, very thin] ([shift={(-3.6cm, -2.4cm)}]schematic.center) grid ([shift={(3.6cm, 1.8cm)}]schematic.center);
    \draw[->, modelgray!60, thick] ([shift={(-3.4cm, -2.2cm)}]schematic.center) -- ([shift={(3.4cm, -2.2cm)}]schematic.center) node[right, font=\Large\sffamily] {};
    \draw[->, modelgray!60, thick] ([shift={(-3.2cm, -2.4cm)}]schematic.center) -- ([shift={(-3.2cm, 1.8cm)}]schematic.center) node[above, font=\Large\sffamily] {};

    \coordinate (muCenter) at ([shift={(-1.0cm, 0.6cm)}]schematic.center);
    \fill[protoviolet!10] (muCenter) ellipse (1.8cm and 0.9cm);
    \fill[protoviolet!25] (muCenter) ellipse (1.2cm and 0.6cm);
    \fill[protoviolet!45] (muCenter) ellipse (0.6cm and 0.3cm);
    \fill[protoviolet] (muCenter) circle (4pt) node[above left, font=\LARGE\sffamily, text=black, align=center] { {\Large(Source)}\\$\boldsymbol{\mu}_E, \boldsymbol{\Sigma}_E$ }; 

    \coordinate (muCenter2) at ([shift={(1.8cm, 0.8cm)}]schematic.center);
    \fill[protoviolet!10] (muCenter2) ellipse (0.9cm and 0.45cm);
    \fill[protoviolet!25] (muCenter2) ellipse (0.5cm and 0.25cm);
    \fill[protoviolet] (muCenter2) circle (3.5pt);

    \coordinate (targetPoint) at ([shift={(1.8cm, -0.6cm)}]schematic.center);
    \fill[tgtgreen] (targetPoint) circle (5pt) node[right, font=\LARGE\sffamily, text=black, align=left, xshift=4pt] {$\mathbf{e}_j$\\ {\Large (Target)}}; 

    \fill[tgtgreen!60] ([shift={(1.2cm, -1.4cm)}]schematic.center) circle (4pt);
    \fill[tgtgreen!60] ([shift={(2.5cm, -1.6cm)}]schematic.center) circle (4pt);
    \fill[tgtgreen!60] ([shift={(2.8cm, -0.4cm)}]schematic.center) circle (4pt);

    \draw[->, alignorange, line width=2.5pt] (targetPoint) -- (muCenter) 
        node[midway, sloped, above, font=\large\bfseries\sffamily, text=black, inner sep=4pt] {Minimize $d_M$}; 
\end{scope}

\draw[darr=modelgray] (matchProto.south) -- ([xshift=-1.2cm]schematic.north);


\draw[arr=modelgray] (UM.north) -- (srcData.south);
\draw[arr=modelgray] (UM.east) -- (tgtData.west);

\draw[arr=srcblue] (srcData.east) -- (extractLatent.west);
\draw[arr=protoviolet] (extractLatent.east) -- (pca.west);
\draw[arr=protoviolet] (pca.east) -- (gmm.west);
\draw[arr=protoviolet] (gmm.east) -- (intProto.west);
\draw[arr=protoviolet] (intProto.east) -- (engProto.west);

\draw[arr=tgtgreen] (tgtData.east) -- (fwdTarget.west);
\draw[arr=tgtgreen] (fwdTarget.east) -- (tgtPca.west) 
    node[midway, above, note, yshift=10pt] {\textbf{latents} \\ $\mathbf{h}_j$};
\draw[arr=tgtgreen] (tgtPca.east) -- (matchProto.west);

\draw[darr=protoviolet] (intProto.south) -- (matchProto.north)
    node[midway, right, note] { $\boldsymbol{\mu}_I, \boldsymbol{\Sigma}_I$};

\draw[arr=alignorange] (matchProto.west) -- ++(-1.2, 0) |- (combinedLoss.east)
    node[pos=0.85, below, note, xshift=15pt] {$\mathcal{L}_{\text{NEAL}}$};

\draw[arr=tgtgreen] (tgtData.south) |- (combinedLoss.west)
    node[near end, below, note, xshift=-5pt] {\textbf{target}\\\textbf{labels}};

\draw[arr=alignorange, line width=2.5pt] (combinedLoss.south) -- (output.north);

\end{tikzpicture}}
    \caption{\small \textbf{Prototype latent alignment workflow}: The pipeline proceeds in three stages to produce a fine-tuned model. Prototype Extraction: Latent representations ($\boldsymbol{h_j}$) from a source subset ($\mathcal{D}_s'$) are projected via PCA and clustered using a Gaussian Mixture Model to obtain source interaction prototypes ($\boldsymbol{\mu_I}, \boldsymbol{\Sigma_I}$). Prototype Assignment: Target samples ($\mathcal{D}_t$) are passed through the foundation model to obtain latents ($h_j$), which are similarly projected, and assigned to the closest interaction prototype via Mahalanobis distance ($d_M$), entirely within the interaction latent space. Prototype Alignment: The assigned distance ($d_M$) is minimized by aligning target nodal energy latents ($\boldsymbol{e_j}$) with the corresponding energy prototypes ($\boldsymbol{\mu_E}, \boldsymbol{\Sigma_E}$). The model is optimized end-to-end using a joint loss, $\mathcal{L}_{\text{Total}} = \mathcal{L}_{\text{Vanilla}} + \lambda_{\text{NEAL}} \mathcal{L}_{\text{NEAL}}$, yielding the final fine-tuned model.}
    \label{fig:neal_workflow}
\end{figure}
\section{Methodology}
Adapting pretrained machine learning interatomic potentials (MLIPs) to new material systems under limited supervision remains a central challenge. While FM  trained on large and diverse datasets provide a strong initialization, their direct fine-tuning on small target datasets often leads to overfitting and poor generalization due to distributional mismatch \cite{liu2025finetuninguniversalmachinelearnedinteratomic}. In this work, we introduce a novel fine-tuning framework that departs from standard target driven regression focusing instead on the explicit alignment of intermediate latent representations. We begin with standard fine-tuning, which we call vanilla fine-tuning, highlight its limitations, and then introduce our prototype-guided alignment method that explicitly enforces consistency between structurally similar atomic environments across different molecular systems. Our problem setting consists of an FM, a subset of the source dataset used to train it and a target system of interest unseen in the source training data.

\subsection{Vanilla Fine-Tuning}

We first consider the standard fine-tuning setting, where an FM is fine-tuned to a target system of interest using a small labelled dataset of molecular configurations of the target molecule. Each configuration consists of the nuclear charges ($z$) of respective atoms, three-dimensional position coordinates ($\textbf{r}$) for the atoms in the target molecules, along with reference energies  ($E$) and (optionally) atomic force labels ($\textbf{F}$) generally obtained from DFT calculations. The training objective minimizes a weighted mean squared error over total energies and atomic forces. 

\begin{equation}
    \mathcal{L}_{\text{Vanilla}}
    =
    \frac{\lambda_E}{m}
    \sum_{i=1}^{m}
    \left| E^{(i)} - \hat{E}^{(i)} \right|^2
    +
    \frac{\lambda_F}{m}
    \sum_{i=1}^{m}
    \frac{1}{3 n^{(i)}}
    \sum_{j=1}^{n^{(i)}}
    \sum_{k=1}^{3}
    \left| \textbf{F}^{(i)}_{jk} - \hat{\textbf{F}}^{(i)}_{jk} \right|^2 .
    \label{eq:loss}
\end{equation}

where:
\begin{itemize}
    \item $\lambda_E$ and $\lambda_F$ are global weighting factors for the energy and force loss, respectively.
    \item $E^{(i)}$ and $\hat{E}^{(i)}$ denote the ground-truth and predicted potential energy for molecule $i$.
    \item $\textbf{F}^{(i)}_{jk}$ and $\hat{\textbf{F}}^{(i)}_{jk}$ denote the ground-truth and predicted force on atom $j$ of molecule $i$ along the $k$-th Cartesian coordinate. 
    \item $n^{(i)}$ is the number of atoms in molecule $i$, and $m$ is the total number of molecules in a minibatch.
\end{itemize}
while simple and widely adopted, this approach assumes that the pretrained representation is already well-aligned with the target distribution. In practice, however, significant distribution shifts arising from differences in chemistry, coordination environments, or thermodynamic conditions lead to inefficient learning.

\FloatBarrier

\begin{figure}[htbp]
    \centering
     \includegraphics[width=0.8\textwidth]{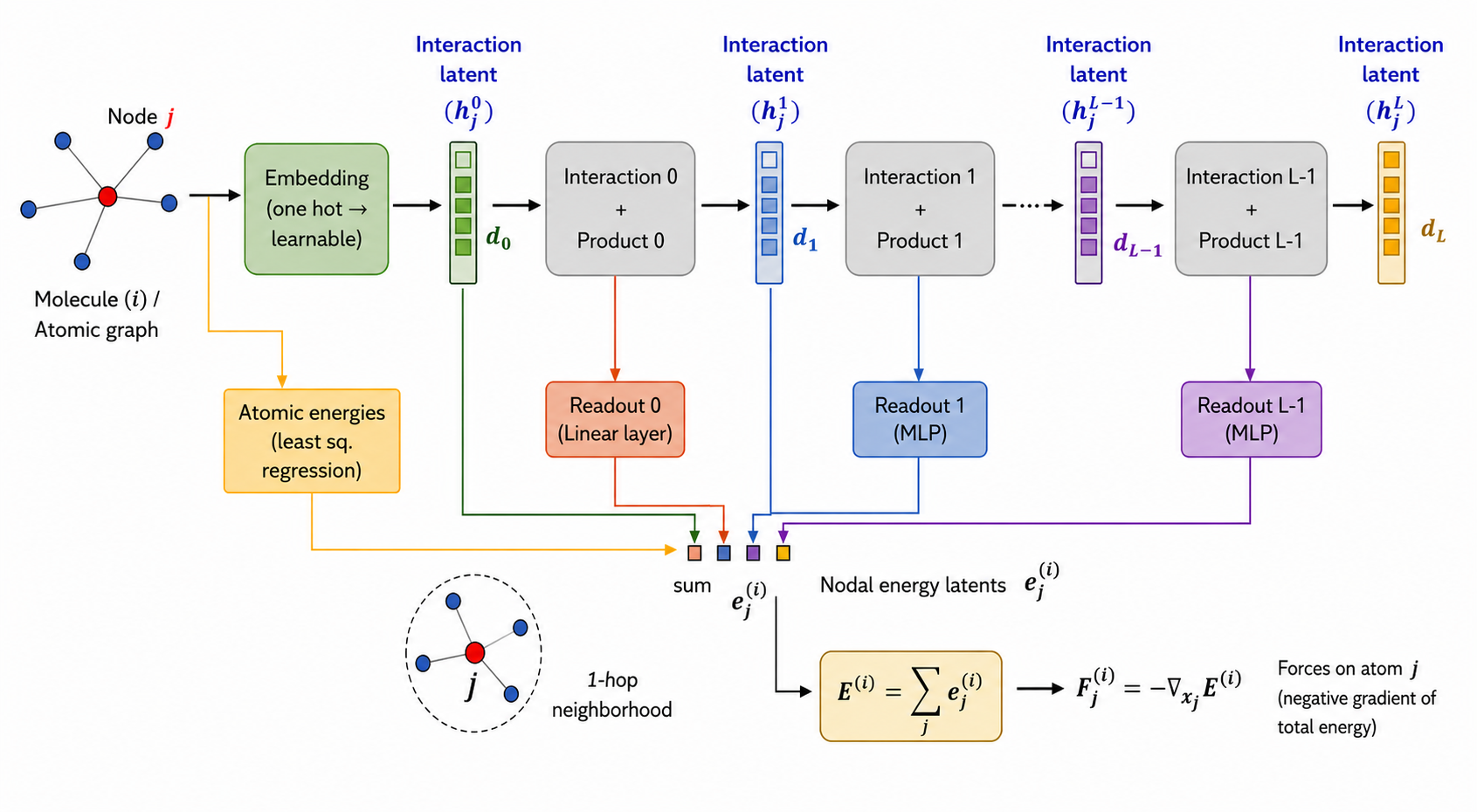}
    \caption{
Architectural overview of MACE with interaction and nodal energy latents. 
For atom $j$ in molecule $i$, its local 1-hop neighborhood is embedded into an initial interaction latent $h_j^{(0)}$. 
This representation is iteratively refined through $L$ interaction blocks, producing updated latents $h_j^{(l)}$. 
At each stage, readout networks map these interaction latents to nodal energy latents. 
These contributions, together with atomic energy terms, are aggregated to obtain the final nodal energy, from which total energy is computed and forces are derived as energy gradients with respect to atomic positions (Figure inspired by \cite{batatia2022mace}). }
    \label{fig:data_graphs}
\end{figure}

\subsection{Prototype Guided Latent Alignment}
To explicitly address the misalignment between learned structural representations of local environments and their corresponding energy contributions across different target systems, we propose a prototype-guided latent alignment framework. We define a \textit{prototype} as a representative of a local atomic environment observed in the source training data. It corresponds to a cluster of similar local environments found in the
source training data. Each cluster is associated with a prototype that has two distinct representations: one in the interaction latent space (interaction prototype) and another in the nodal energy contribution space (energy prototype). For each atom $j$,  most of the GNN architectures  produces an interaction latent vector $\mathbf{h}_{j}$, which encodes the local atomic environment through message passing, capturing both geometric and chemical context within a finite neighborhood. This representation serves as a structural descriptor of the atom's environment. In addition, the model computes a nodal energy contribution representation $\mathbf{e}_{j}$, which is derived from $\mathbf{h}_{j}$ and encodes the atom-wise contribution to the total energy.
\subsubsection{Prototype Extraction}
We begin by extracting prototypes from a  source training data. To systematically extract these interactions and energy prototypes, we utilize a subset of the source training data. This subset ($\mathcal{D^{\prime}_{\text{s}}}$) is constructed via compound-wise random stratified sampling without replacement and a predetermined per-compound budget. This ensures that our extracted prototypes effectively characterize the entire pre-training dataset. We collect the interaction latent representations produced by the FM for the atoms in this subset and apply principal component analysis (PCA) \cite{abdi2010principal} for dimensionality reduction. This dimensionality reduction reduces the computational burden and mitigates the curse of dimensionality.

This is followed by clustering for each compound, element pair using Gaussian mixture models (GMMs) \cite{reynolds2009gaussian}, where the resulting cluster representatives ($\boldsymbol{\mu}_{I}, \boldsymbol{\Sigma}_{I}$) define the set of interaction prototypes while using Bayesian Information Criterion (BIC) \cite{vrieze2012model} score to determine the optimal number of clusters.  Subsequently, we construct the corresponding energy prototypes by modeling the distribution of nodal energy contributions ($\boldsymbol{e_{j}}$) within each cluster using a Gaussian distribution. Specifically, for the set of atoms assigned to a given interaction cluster, we fit a Gaussian probability density function and estimate its parameters—the mean ($\boldsymbol{\mu_E}$) and covariance ($\boldsymbol{\Sigma_E}$). These parameters serve as the energy prototypes associated with each interaction prototype. The extraction procedure is outlined in algorithms \ref{alg:pca_gmm_prototypes} and \ref{alg:energy_prototype_extraction} in the appendix.

\subsubsection{Prototype Assignment}
Once the prototypes are defined, each atom in the target system is assigned to its closest interaction prototype. The assignment of atoms in the target system to prototypes is performed in the interaction latent space. We first extract the interaction latent vectors ($\boldsymbol{h_{j}}$) from the FM for each atom $j$ in the target system. These vectors are then projected into the
reduced PCA space learned from the subset of source training data. Given the set of Gaussian clusters  $\{(\boldsymbol{\mu}_I^{(k)}, \boldsymbol{\Sigma}_I^{(k)})\}$, the assignment for a target atom $j$ with projected latent $\boldsymbol{\tilde{h}_j}$  is made by finding the cluster with the minimum Mahalanobis distance which is a natural choice for distance metric since we assumed a Gaussian distribution.
\begin{equation}
k_j^* = \arg\min_{k} \, d_M\!\left(\tilde{\mathbf{h}}_j, \boldsymbol{\mu}_I^{(k)}, \boldsymbol{\Sigma}_I^{(k)}\right)
= \arg\min_{k}
\sqrt{
\left(\tilde{\mathbf{h}}_j - \boldsymbol{\mu}_I^{(k)}\right)^{\top}
\left(\boldsymbol{\Sigma}_I^{(k)}\right)^{-1}
\left(\tilde{\mathbf{h}}_j - \boldsymbol{\mu}_I^{(k)}\right)
}.
\end{equation}

Once each atom j is assigned to a cluster $k^{*}_{j}$, the corresponding energy prototype parameters  $\left(\boldsymbol{\mu}_E^{(k_j^*)}, \boldsymbol{\Sigma}_E^{(k_j^*)}\right)$, learned from the source training data cluster's nodal energy contribution latent vectors are used for alignment.

\subsubsection{Prototype Alignment}
To enforce  consistency between target nodal energy latents and prototype statistics, we introduce the Nodal Energy Alignment Loss (NEAL). NEAL encourages the nodal energy contribution latent vector  $\boldsymbol{e_{j}}$  of an atom $j$ in a molecule from the target system to align closely with its assigned nodal energy contribution prototype. For a molecule with $n$ atoms we have

\begin{equation}
    \mathcal{L}_{\text{NEAL}} = \frac{1}{n}\sum_{j=1}^{j=n}\left(e_j-\boldsymbol{\mu}^{k^{*}_{j}}_{E}\right)^{\top}(\boldsymbol{\Sigma}^{k^{*}_{j}}_{E})^{-1}\left(e_j-\boldsymbol{\mu}^{k^{*}_{j}}_{E}\right)
\end{equation}

We apply this additional alignment objective together with the vanilla loss to enforce consistency between the energy labels of structurally similar samples in the source and target systems. However, applying this constraint throughout the entire fine-tuning process can introduce excessive bias toward the source-system representations and hinder learning the target system nuances. Therefore, we apply the alignment only during the initial epochs of fine-tuning. Refer to Appendix~\ref{app:effect_of_n_align_and_lambda_neal} for additional details.

This can be thought of as adding an additional bias term to model fine-tuning common in low data settings to reduce the effect of variance in over-parametrized models. This also provides a per atom energy guidance to the model unlike the vanilla loss where an overall guidance on the total energy of the configuration is provided. The total loss at epoch $n_{e}$ is given by
\begin{equation}
    \mathcal{L}_{\text{Total}} =
    \begin{cases}
        \mathcal{L}_{\text{Vanilla}} + \lambda_{\text{NEAL}} \mathcal{L}_{\text{NEAL}} & \text{if } n_{e} \le N_{\text{NEAL}} \\
        \mathcal{L}_{\text{Vanilla}}                                                    & \text{if } n_{e} > N_{\text{NEAL}}
    \end{cases}
    \label{eq:neal_loss}
\end{equation}
where $N_{\text{NEAL}}$ is the epoch threshold after which the alignment loss is deactivated. This introduces two additional hyperparameters $\lambda_{\text{NEAL}}$ and $N_{\text{NEAL}}$ that can be tuned to optimize the performance of the model on the target system. Removing this alignment loss for the later epochs ensures that the model is able to learn the target system specific interactions that may not be present in the source training data. This can be thought of as guiding the model with prior knowledge from the source training data and later allowing it to reach a better minima when this guidance is removed than possible in the absence of the alignment based guidance.

\section{Experimental Setup}
In this section, we detail the experimental procedures used to evaluate our proposed Neal alignment framework. First, we describe our source-target transfer setup using proxy foundation models on the rMD17 dataset. Then, we outline our evaluation on real, large-scale foundation models using the MACEOFF series.

\subsection{Experiments with Proxy Foundation Models}

Our proxy foundation model is first trained on source training data using joint supervision on total energy and atomic forces via the weighted MSE objective in Eq.~\ref{eq:loss}. We then fine-tune this pretrained model on each target molecule separately using the same energy-force objective, using both vanilla and prototype latent alignment framework on SchNet and MACE architectures.

We use the rMD17 dataset \citep{Christensen2020}, which provides molecular dynamics trajectories for ten organic molecules: aspirin, azobenzene, benzene, ethanol, malonaldehyde, naphthalene, paracetamol, salicylic acid, toluene, and uracil. Each system is annotated with potential energies and atomic forces.

To construct source-target transfer settings with varying levels of difficulty, we define two scenarios. In the first setting (ST1), all local chemical environments present in the target system are also represented in the source training data. In the second setting (ST2), we increase the transfer difficulty by placing azobenzene in the target dataset, introducing an \(\mathrm{N{=}N}\) functional group that is absent from the source dataset.

\begin{figure}[h!]
    \centering
    \includegraphics[width=\linewidth]{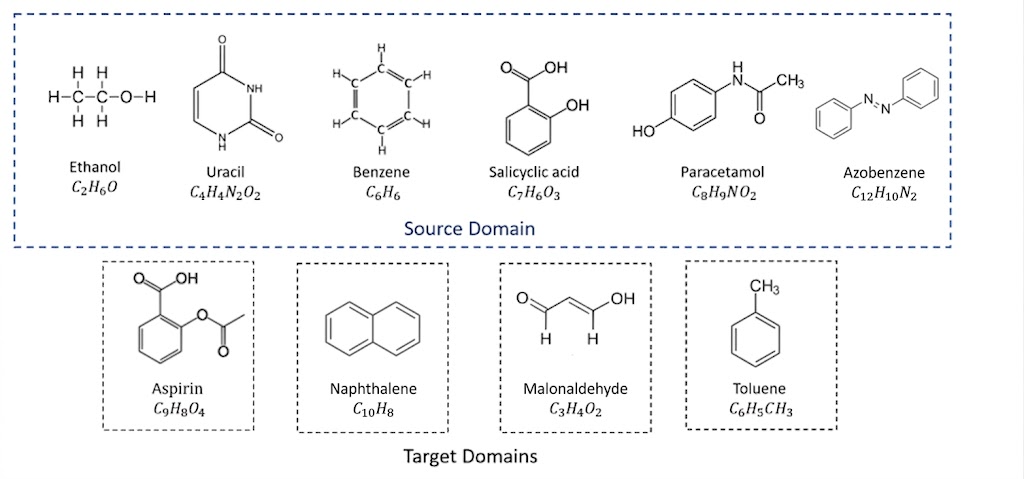}
    \caption{Source and target molecules used in the rMD17 transfer setup in ST1.}
    \label{fig:domain_split}
\end{figure}

Figure~\ref{fig:domain_split} visualizes the ST1 partition.

\newpage

\begin{itemize}
    \item \textbf{ST1}
    \begin{itemize}
        \item \textbf{Source dataset:} ethanol, uracil, benzene, salicylic acid, azobenzene, and paracetamol.
        \item \textbf{Target dataset:} aspirin, naphthalene, malonaldehyde, and toluene.
    \end{itemize}

    \item \textbf{ST2}
    \begin{itemize}
        \item \textbf{Source dataset:} ethanol, uracil, benzene, salicylic acid, naphthalene, and paracetamol.
        \item \textbf{Target dataset:} aspirin, azobenzene, malonaldehyde, and toluene.
    \end{itemize}
\end{itemize}

\subsection{Experiments with Foundation Models}

To evaluate our approach on real Foundation models, we conduct additional experiments using the MACEOFF23 series, a family of high-fidelity interatomic potentials for organic molecules trained primarily on version~1 of the SPICE dataset~\citep{eastman_2022_8222043}. We select target systems from SPICE version~2.0.1~\citep{eastman_2024_10975225}. To avoid data leakage, we restrict evaluation to molecules whose SMILES strings are present only in the newer SPICE release and absent from the earlier version used to train the MACEOFF23 models.

\section{Main Results}
All energy-MAE values are reported in meV, while all force-MAE values are reported in meV/\AA{}. Reported metrics are averaged across multiple random seeds. Additional hyperparameter details are provided in Appendix~\ref{hyperparameter-details}. The lowest energy- and force-MAE values in each row are highlighted in bold. Statistically significant values at 95\% significance, calculated via the Wilcoxon signed-rank test, are marked by *.
\subsection{SchNet Results}

\textbf{ST1.}\enspace Table~\ref{tab:schnet_st1_results} reports SchNet performance on ST1, where all local chemical environments in the target molecules are represented in the source training data. NEAL consistently improves energy MAE over the vanilla baseline across all four target molecules and all training-set sizes. The gains are largest in the most data-scarce regime: for Aspirin at 5 samples, NEAL reduces energy MAE from 240.1 to 235.7~meV; for Malonaldehyde at the same size, the reduction is from 113.1 to 97.6~meV ($\approx$14\%). Energy improvements remain meaningful even at 40 samples, where Malonaldehyde benefits by $\approx$13~meV. Force MAE improvements are present but more variable: NEAL leads on Malonaldehyde and Naphthalene across most sizes, while results for Aspirin and Toluene are mixed, with neither method dominating consistently. The overall picture is that prototype-guided alignment provides a clear, reliable benefit for energy prediction on ST1 and a moderate, less uniform benefit for forces.

\textbf{ST2.}\enspace Table~\ref{tab:schnet_st2_results} evaluates the harder transfer setting in which azobenzene-whose N=N double bond is absent from source data-is included as a target. For Aspirin, NEAL achieves substantial energy MAE gains of up to $\approx$15~meV at 40 samples, and force MAE gains at 10 and 20 samples. For Malonaldehyde and Toluene, NEAL consistently leads on energy at 10 samples and above. Azobenzene is the exception: vanilla matches or narrowly outperforms NEAL on energy at 5 and 40 samples, while force gains are marginal. This irregular pattern is directly attributable to structural mismatch in the assigned prototypes for nitrogen atoms.

\textbf{Azobenzene ablation.}\enspace To isolate the effect of prototype misalignment, we ablate the matching loss for nitrogen atoms in azobenzene and re-evaluate (Table~\ref{tab:azobenzene_alignment}). With nitrogen alignment suppressed, NEAL (w/o N) outperforms vanilla at every training-set size in both energy and force MAE. At 5 samples, energy MAE drops from 106.7 to 90.2~meV and force MAE from 179.5 to 166.7~meV/\AA. These improvements are consistent across 10, 20, and 40 samples. This result confirms that the failure on azobenzene in ST2 is entirely attributable to the misaligned N prototype, not to any weakness of the alignment mechanism per se.

This sensitivity to prototype validity is itself informative. Unlike conventional regularization, which applies data-agnostic, uniform constraints, NEAL's structured guidance actively penalizes correspondences that are structurally invalid. Misaligned guidance therefore exposes a failure mode invisible to standard regularization schemes. In the low-data proxy setting this requires per-atom-type attention; in large-scale foundation models trained on chemically diverse data, structurally invalid assignments become rare, and the mechanism instead acts as a precise filter enforcing accurate, structure-aware inductive bias across a broad chemical space.

\begin{table}[ht]
    \centering

    \begin{minipage}[t]{0.49\textwidth}
        \centering
        \caption{SchNet results on ST1.}
        \label{tab:schnet_st1_results}
        \vspace{8pt}
        \small
        \setlength{\tabcolsep}{2.5pt}
        \resizebox{\linewidth}{!}{%
            \begin{tabular}{llcccc}
                \toprule
                \textbf{Molecule} & \textbf{\#Ex} & \multicolumn{2}{c}{\textbf{Energy MAE}$\downarrow$} & \multicolumn{2}{c}{\textbf{Force MAE}$\downarrow$}                                                                                           \\
                \cmidrule(lr){3-4}\cmidrule(lr){5-6}
                                  &               & \textbf{NEAL}                                       & \textbf{Vanilla}                                   & \textbf{NEAL}               & \textbf{Vanilla}            \\
                \midrule
                \multirow{4}{*}{Aspirin}
                                  & 5             & \textbf{218.95 $\pm$ 30.90*}                        & 227.63 $\pm$ 21.81                                 & 299.36 $\pm$ 13.28          & \textbf{297.98 $\pm$ 14.76} \\
                                  & 10            & \textbf{138.30 $\pm$ 9.23*}                         & 157.54 $\pm$ 11.77                                 & 251.25 $\pm$ 6.49           & \textbf{248.10 $\pm$ 10.59} \\
                                  & 20            & \textbf{104.30 $\pm$ 6.66*}                         & 114.70 $\pm$ 7.81                                  & \textbf{203.52 $\pm$ 4.18}  & 205.42 $\pm$ 7.31           \\
                                  & 40            & \textbf{67.79 $\pm$ 4.70*}                          & 72.80 $\pm$ 5.14                                   & 152.36 $\pm$ 2.89           & \textbf{151.23 $\pm$ 2.84}  \\
                \midrule
                \multirow{4}{*}{Malonaldehyde}
                                  & 5             & \textbf{115.02 $\pm$ 25.82*}                        & 138.68 $\pm$ 36.31                                 & \textbf{335.95 $\pm$ 11.13} & 344.68 $\pm$ 19.28          \\
                                  & 10            & \textbf{91.29 $\pm$ 11.85*}                         & 115.71 $\pm$ 28.15                                 & \textbf{248.94 $\pm$ 6.37*} & 251.94 $\pm$ 11.41          \\
                                  & 20            & \textbf{56.47 $\pm$ 4.75*}                          & 72.73 $\pm$ 8.45                                   & \textbf{178.32 $\pm$ 9.02}  & 180.46 $\pm$ 7.69           \\
                                  & 40            & \textbf{39.35 $\pm$ 3.08*}                          & 55.05 $\pm$ 5.73                                   & \textbf{136.04 $\pm$ 4.54}  & 138.87 $\pm$ 3.85           \\
                \midrule
                \multirow{4}{*}{Naphthalene}
                                  & 5             & \textbf{36.10 $\pm$ 5.83*}                          & 47.62 $\pm$ 17.73                                  & 104.40 $\pm$ 3.57           & \textbf{103.91 $\pm$ 4.54}  \\
                                  & 10            & \textbf{23.22 $\pm$ 1.94*}                          & 25.55 $\pm$ 2.03                                   & \textbf{77.54 $\pm$ 2.81}   & 78.80 $\pm$ 3.06            \\
                                  & 20            & \textbf{16.99 $\pm$ 1.22*}                          & 17.41 $\pm$ 0.74                                   & 59.62 $\pm$ 1.26            & \textbf{59.36 $\pm$ 1.29}   \\
                                  & 40            & \textbf{13.67 $\pm$ 0.79*}                          & 14.78 $\pm$ 1.24                                   & \textbf{45.49 $\pm$ 0.50}   & 48.39 $\pm$ 1.49            \\
                \midrule
                \multirow{4}{*}{Toluene}
                                  & 5             & \textbf{41.29 $\pm$ 6.32}                           & 43.65 $\pm$ 9.22                                   & 129.26 $\pm$ 7.15           & \textbf{128.86 $\pm$ 7.37}  \\
                                  & 10            & \textbf{28.42 $\pm$ 2.65}                           & 32.84 $\pm$ 5.63                                   & 97.01 $\pm$ 7.60            & \textbf{95.44 $\pm$ 5.62}   \\
                                  & 20            & \textbf{21.58 $\pm$ 3.09*}                          & 22.03 $\pm$ 3.09                                   & 72.42 $\pm$ 3.09            & \textbf{72.38 $\pm$ 2.66}   \\
                                  & 40            & \textbf{15.66 $\pm$ 1.24}                           & 15.68 $\pm$ 1.66                                   & \textbf{56.28 $\pm$ 1.15}   & 56.63 $\pm$ 1.46            \\
                \bottomrule
            \end{tabular}%
        }
    \end{minipage}\hfill
\begin{minipage}[t]{0.49\textwidth}
        \centering
        \caption{SchNet results on ST2.}
        \label{tab:schnet_st2_results}
        \vspace{8pt}
        \small
        \setlength{\tabcolsep}{2.5pt}
        \resizebox{\linewidth}{!}{%
            \begin{tabular}{llcccc}
                \toprule
                \textbf{Molecule} & \textbf{\#Ex} & \multicolumn{2}{c}{\textbf{Energy MAE}$\downarrow$} & \multicolumn{2}{c}{\textbf{Force MAE}$\downarrow$}                                \\
                \cmidrule(lr){3-4}\cmidrule(lr){5-6}
                                  &               & \textbf{NEAL}                                       & \textbf{Vanilla}                                   & \textbf{NEAL}               & \textbf{Vanilla}            \\
                \midrule
                \multirow{4}{*}{Aspirin}
                                  & 5             & 263.19 $\pm$ 35.52                                  & \textbf{254.81 $\pm$ 39.76}                        & 294.66 $\pm$ 19.63          & \textbf{289.37 $\pm$ 16.08} \\
                                  & 10            & \textbf{174.07 $\pm$ 26.55}                         & 180.64 $\pm$ 20.52                                 & \textbf{223.58 $\pm$ 9.99}  & 225.99 $\pm$ 6.40           \\
                                  & 20            & \textbf{111.29 $\pm$ 18.47}*                        & 131.59 $\pm$ 7.77                                  & 178.66 $\pm$ 7.00           & \textbf{173.97 $\pm$ 4.24}  \\
                                  & 40            & \textbf{72.71 $\pm$ 5.87}*                          & 87.27 $\pm$ 16.59                                  & 138.86 $\pm$ 3.27           & \textbf{138.52 $\pm$ 2.94}  \\
                \midrule
                \multirow{4}{*}{Azobenzene}
                                  & 5             & \textbf{92.82 $\pm$ 17.88}*                         & 106.67 $\pm$ 15.09                                 & \textbf{169.99 $\pm$ 12.32}* & 179.49 $\pm$ 13.75          \\
                                  & 10            & 70.56 $\pm$ 3.03*                                   & \textbf{70.03 $\pm$ 7.35}                          & \textbf{145.25 $\pm$ 9.21}  & 145.29 $\pm$ 4.32           \\
                                  & 20            & \textbf{49.47 $\pm$ 8.16}*                          & 52.16 $\pm$ 9.82                                   & \textbf{109.66 $\pm$ 6.08}* & 115.26 $\pm$ 2.21           \\
                                  & 40            & 38.76 $\pm$ 4.35*                                   & \textbf{38.27 $\pm$ 2.36}                          & \textbf{85.99 $\pm$ 5.39}   & 89.45 $\pm$ 2.30            \\
                \midrule
                \multirow{4}{*}{Malonaldehyde}
                                  & 5             & 113.89 $\pm$ 20.76                                  & \textbf{110.69 $\pm$ 19.78}                        & 299.02 $\pm$ 7.99           & \textbf{296.45 $\pm$ 12.54} \\
                                  & 10            & \textbf{84.06 $\pm$ 12.83}*                         & 92.36 $\pm$ 14.32                                  & 231.28 $\pm$ 9.36           & \textbf{229.65 $\pm$ 15.20} \\
                                  & 20            & \textbf{67.60 $\pm$ 11.35}*                         & 72.20 $\pm$ 8.46                                   & 173.35 $\pm$ 2.64           & \textbf{173.30 $\pm$ 4.42}  \\
                                  & 40            & \textbf{40.13 $\pm$ 3.84}*                          & 44.51 $\pm$ 6.27                                   & 135.32 $\pm$ 5.28*          & \textbf{134.94 $\pm$ 3.82}  \\
                \midrule
                \multirow{4}{*}{Toluene}
                                  & 5             & \textbf{33.75 $\pm$ 6.23}                           & 33.83 $\pm$ 7.47                                   & 106.12 $\pm$ 3.75           & \textbf{101.91 $\pm$ 2.51}  \\
                                  & 10            & \textbf{22.89 $\pm$ 2.08}                           & 24.74 $\pm$ 5.24                                   & 82.50 $\pm$ 3.25            & \textbf{81.00 $\pm$ 3.83}   \\
                                  & 20            & \textbf{17.43 $\pm$ 0.96}*                          & 17.53 $\pm$ 0.84                                   & \textbf{65.67 $\pm$ 1.71}   & 66.32 $\pm$ 2.04            \\
                                  & 40            & \textbf{12.57 $\pm$ 0.68}                           & 13.27 $\pm$ 1.68                                   & \textbf{47.31 $\pm$ 1.05}*  & 49.74 $\pm$ 2.09            \\
                \bottomrule
            \end{tabular}%
        }
    \end{minipage}
\end{table}

\begin{table}[ht]
    \centering

    \begin{minipage}[t]{0.49\textwidth}
        \centering
        \caption{Azobenzene with alignment for nitrogen removed.}
        \label{tab:azobenzene_alignment}
        \vspace{8pt}
        \small
        \setlength{\tabcolsep}{2.5pt}
        \resizebox{\linewidth}{!}{%
            \begin{tabular}{llcccc}
                \toprule
                \textbf{Molecule} & \textbf{\#Ex} & \multicolumn{2}{c}{\textbf{Energy MAE} $\downarrow$} & \multicolumn{2}{c}{\textbf{Force MAE} $\downarrow$}                                                    \\
                \cmidrule(lr){3-4} \cmidrule(lr){5-6}
                                  &               & \textbf{NEAL (w/o N)}                                & \textbf{Vanilla}                                    & \textbf{NEAL (w/o N)}       & \textbf{Vanilla}   \\
                \midrule
                \multirow{4}{*}{Azobenzene}
                                  & 5             & \textbf{90.17 $\pm$ 10.12}                           & 106.67 $\pm$ 15.08                                  & \textbf{166.72 $\pm$ 17.54} & 179.49 $\pm$ 13.75 \\
                                  & 10            & \textbf{68.27 $\pm$ 4.43}                            & 70.03 $\pm$ 7.35                                    & \textbf{141.66 $\pm$ 13.52} & 145.29 $\pm$ 4.32  \\
                                  & 20            & \textbf{47.86 $\pm$ 6.08}                            & 52.16 $\pm$ 9.81                                    & \textbf{107.06 $\pm$ 9.67}  & 115.26 $\pm$ 2.21  \\
                                  & 40            & \textbf{37.34 $\pm$ 3.50}                            & 38.27 $\pm$ 2.36                                    & \textbf{82.08 $\pm$ 7.87}   & 89.45 $\pm$ 2.30   \\
                \bottomrule
            \end{tabular}%
        }
    \end{minipage}\hfill
    \begin{minipage}[t]{0.49\textwidth}
        \centering
        \caption{Combined results on various compounds using MACEOFF23 across different numbers of training examples.}
        \label{tab:combined_results}
        \vspace{8pt}
        \small
        \setlength{\tabcolsep}{2.5pt}
        \resizebox{\linewidth}{!}{%
            \begin{tabular}{llcccc}
                \toprule
                \textbf{Compound} & \textbf{\#Ex} & \multicolumn{2}{c}{\textbf{Energy MAE} $\downarrow$} & \multicolumn{2}{c}{\textbf{Force MAE} $\downarrow$}                                                         \\
                \cmidrule(lr){3-4} \cmidrule(lr){5-6}
                                  &               & \textbf{NEAL}                                        & \textbf{Vanilla}                                    & \textbf{NEAL}             & \textbf{Vanilla}          \\
                \midrule
                \multirow{2}{*}{104069541}
                                  & 5             & \textbf{11.13 $\pm$ 3.82}                            & 11.26 $\pm$ 3.72                                    & \textbf{46.20 $\pm$ 3.67} & 46.24 $\pm$ 3.55          \\
                                  & 10            & 8.04 $\pm$ 1.17                                      & \textbf{8.01 $\pm$ 1.45}                            & \textbf{34.09 $\pm$ 1.47} & 34.16 $\pm$ 1.35          \\
                \midrule
                \multirow{2}{*}{135030663}
                                  & 5             & \textbf{34.62 $\pm$ 6.62}                            & 34.65 $\pm$ 6.58                                    & \textbf{56.25 $\pm$ 1.52} & 56.34 $\pm$ 1.55          \\
                                  & 10            & 21.79 $\pm$ 3.39                                     & \textbf{21.50 $\pm$ 3.82}                           & \textbf{48.89 $\pm$ 2.43} & 48.90 $\pm$ 2.44          \\
                \midrule
                \multirow{2}{*}{135094581}
                                  & 5             & 6.54 $\pm$ 0.42                                      & \textbf{6.52 $\pm$ 0.35}                            & \textbf{35.34 $\pm$ 1.49} & 35.44 $\pm$ 1.51          \\
                                  & 10            & \textbf{5.32 $\pm$ 0.87}                             & 5.34 $\pm$ 0.98                                     & 28.02 $\pm$ 1.68          & \textbf{28.02 $\pm$ 1.60} \\
                \midrule
                \multirow{2}{*}{135161703}
                                  & 5             & \textbf{25.46 $\pm$ 4.45}                            & 25.21 $\pm$ 4.56                                    & \textbf{81.92 $\pm$ 1.35} & 81.92 $\pm$ 1.35          \\
                                  & 10            & \textbf{24.22 $\pm$ 2.72}                            & 24.93 $\pm$ 3.13                                    & \textbf{74.18 $\pm$ 1.64} & 74.33 $\pm$ 2.01          \\
                \midrule
                \multirow{2}{*}{135170767}
                                  & 5             & \textbf{17.06 $\pm$ 2.74}                            & 20.12 $\pm$ 1.22                                    & \textbf{34.79 $\pm$ 0.61} & 34.03 $\pm$ 1.57          \\
                                  & 10            & 14.91 $\pm$ 1.02                                     & \textbf{14.71 $\pm$ 1.14}                           & 28.43 $\pm$ 0.67          & \textbf{28.46 $\pm$ 0.66} \\
                \midrule
                \multirow{2}{*}{135246570}
                                  & 5             & \textbf{12.00 $\pm$ 2.23}                            & 12.17 $\pm$ 2.22                                    & \textbf{45.59 $\pm$ 1.66} & 45.76 $\pm$ 1.76          \\
                                  & 10            & \textbf{11.12 $\pm$ 1.21}                            & 11.12 $\pm$ 1.18                                    & 36.48 $\pm$ 2.42          & \textbf{36.48 $\pm$ 2.44} \\
                \bottomrule
            \end{tabular}%
        }
    \end{minipage}

\end{table}

\subsubsection{MACE Results}

\textbf{ST1.}\enspace Table~\ref{tab:mace_st1_results} reports MACE performance on ST1. Prototype-guidance consistently reduces energy MAE relative to the vanilla baseline across all four target molecules and all training-set sizes (5, 10, and 20 samples). The improvements are most pronounced in the lowest-data regime: for Aspirin at 5 samples, NEAL reduces energy MAE from 59.5 to 51.1~meV ($\approx$14\%); for Malonaldehyde the reduction is from 52.3 to 46.5~meV ($\approx$11\%). Gains persist at 20 samples, though they narrow, consistent with greater baseline data sufficiency. Force MAE improvements are present but smaller in absolute terms and occasionally reversed for Toluene at 5 and 10 samples, where vanilla holds a marginal edge.

\textbf{ST2.}\enspace Table~\ref{tab:mace_st2_results_clean} reports MACE performance on the harder ST2 setting. Prototype-guidance yields significant energy MAE improvements for Aspirin ($\approx$9~meV at 5 samples, $\approx$3~meV at 10 samples) and for Malonaldehyde ($\approx$5~meV at 5 and 10 samples). For Azobenzene, NEAL FT achieves significant energy gain at 5 samples ($\approx$2~meV) and leads at 20 samples, despite the N=N structural mismatch that caused difficulties for SchNet. The stronger MACE backbone appears more robust to prototype misalignment, likely because its more expressive equivariant representations provide a richer prior that partially compensates for imperfect guidance. For Toluene, the two methods are effectively equivalent across all sizes, with differences within noise. Force MAE trends mirror the energy results: NEAL FT consistently matches or outperforms VFT for Aspirin and Malonaldehyde, while Toluene and Azobenzene forces show no reliable winner. Overall, these results demonstrate that prototype-guidance enhances both accuracy and data efficiency of MACE-based MLIPs across diverse transfer scenarios.

\textbf{MACEOFF23.}\enspace Table~\ref{tab:combined_results} extends this evaluation to seven additional compounds fine-tuned from MACEOFF23 in the ultra-low-data regime (5 and 10 examples); SMILES strings for all compounds are provided in Appendix~\ref{appendix:smiles}. Across this broader chemical space, NEAL and vanilla fine-tuning are largely equivalent: differences fall within statistical noise for the majority of compound-size pairs, reflecting the strong inductive bias already encoded in the MACEOFF23 pre-trained representations. Nevertheless, NEAL provides measurable energy MAE improvements on several compounds at 5 examples, most notably compound 135170767, where it reduces energy MAE from 20.12 to 17.06~meV and maintains practical parity or better on force MAE in nearly every case. Critically, NEAL never exhibits practical degradation relative to vanilla fine-tuning across any compound or training-set size, establishing it as a generally superior fine-tuning strategy: it matches vanilla in the best case and meaningfully outperforms it when the prototype correspondences can meaningfully contribute towards improvement.

\begin{table}[ht]
    \centering

\begin{minipage}[t]{0.49\textwidth}
        \centering
        \caption{MACE results on ST1.}
        \label{tab:mace_st1_results}
        \vspace{8pt} 
        \small
        \setlength{\tabcolsep}{2.5pt} 
        \resizebox{\linewidth}{!}{%
            \begin{tabular}{llcccc}
                \toprule
                \textbf{Molecule} & \textbf{\#Ex} & \multicolumn{2}{c}{\textbf{Energy MAE}$\downarrow$} & \multicolumn{2}{c}{\textbf{Force MAE}$\downarrow$} \\
                \cmidrule(lr){3-4}\cmidrule(lr){5-6}
                                  &                & \begin{tabular}{@{}c@{}}\textbf{NEAL}\\\vphantom{\textbf{Baseline}}\end{tabular}
                                  & \begin{tabular}{@{}c@{}}\textbf{Vanilla}\\\vphantom{\textbf{Baseline}}\end{tabular}
                                  & \begin{tabular}{@{}c@{}}\textbf{NEAL}\\\vphantom{\textbf{Baseline}}\end{tabular}
                                  & \begin{tabular}{@{}c@{}}\textbf{Vanilla}\\\vphantom{\textbf{Baseline}}\end{tabular} \\
                \midrule
                \multirow{3}{*}{Aspirin}
                                  & 5              & \textbf{47.68 $\pm$ 4.81}*                                                       & 54.53 $\pm$ 7.48                                                   & \textbf{88.50 $\pm$ 3.78}  & 89.02 $\pm$ 6.46            \\
                                  & 10             & \textbf{31.14 $\pm$ 1.81}                                                        & 32.64 $\pm$ 1.87                                                   & 63.50 $\pm$ 3.32            & \textbf{63.35 $\pm$ 3.41} \\
                                  & 20             & 20.24 $\pm$ 1.01                                                                 & \textbf{20.22 $\pm$ 1.14}                                          & \textbf{47.23 $\pm$ 1.41}  & 47.51 $\pm$ 1.47            \\
                \midrule
                \multirow{3}{*}{Malonaldehyde}
                                  & 5              & \textbf{44.46 $\pm$ 4.25}*                                                       & 51.56 $\pm$ 5.91                                                   & \textbf{114.43 $\pm$ 5.08} & 115.85 $\pm$ 3.77          \\
                                  & 10             & \textbf{23.84 $\pm$ 2.38}*                                                       & 26.49 $\pm$ 2.16                                                   & \textbf{79.79 $\pm$ 4.04}*  & 81.66 $\pm$ 4.27            \\
                                  & 20             & \textbf{13.87 $\pm$ 1.09}*                                                       & 14.68 $\pm$ 1.03                                                   & \textbf{57.09 $\pm$ 1.80}*  & 58.23 $\pm$ 1.87            \\
                \midrule
                \multirow{3}{*}{Naphthalene}
                                  & 5              & \textbf{5.85 $\pm$ 2.30}                                                         & 5.94 $\pm$ 2.38                                                    & \textbf{21.92 $\pm$ 5.73}*  & 22.06 $\pm$ 5.64            \\
                                  & 10             & \textbf{3.31 $\pm$ 0.45}                                                         & 3.32 $\pm$ 0.35                                                    & \textbf{14.32 $\pm$ 2.13}*  & 14.51 $\pm$ 2.15            \\
                                  & 20             & 2.24 $\pm$ 0.26                                                                  & \textbf{2.22 $\pm$ 0.26}                                           & \textbf{9.22 $\pm$ 0.58}    & 9.30 $\pm$ 0.53             \\
                \midrule
                \multirow{3}{*}{Toluene}
                                  & 5              & \textbf{7.32 $\pm$ 2.41}                                                         & 7.47 $\pm$ 2.44                                                    & 18.45 $\pm$ 0.73            & \textbf{18.23 $\pm$ 0.71} \\
                                  & 10             & \textbf{3.71 $\pm$ 0.24}                                                         & 3.83 $\pm$ 0.35                                                    & 13.33 $\pm$ 0.47            & \textbf{13.09 $\pm$ 0.51} \\
                                  & 20             & \textbf{2.20 $\pm$ 0.04}                                                         & 2.26 $\pm$ 0.11                                                    & 9.06 $\pm$ 0.09            & \textbf{8.93 $\pm$ 0.09}   \\
                \bottomrule
            \end{tabular}%
        }
    \end{minipage}\hfill
    \begin{minipage}[t]{0.49\textwidth}
        \centering
        \caption{MACE results on ST2.}
        \label{tab:mace_st2_results_clean}
        \vspace{8pt} 
        \small
        \setlength{\tabcolsep}{2.5pt} 
        \resizebox{\linewidth}{!}{%
            \begin{tabular}{llcccc}
                \toprule
                \textbf{Molecule} & \textbf{\#Ex} & \multicolumn{2}{c}{\textbf{Energy MAE}$\downarrow$} & \multicolumn{2}{c}{\textbf{Force MAE}$\downarrow$} \\
                \cmidrule(lr){3-4}\cmidrule(lr){5-6}
                                  &                & \begin{tabular}{@{}c@{}}\textbf{NEAL}\\\vphantom{\textbf{Baseline}}\end{tabular}
                                  & \begin{tabular}{@{}c@{}}\textbf{Vanilla}\\\vphantom{\textbf{Baseline}}\end{tabular}
                                  & \begin{tabular}{@{}c@{}}\textbf{NEAL}\\\vphantom{\textbf{Baseline}}\end{tabular}
                                  & \begin{tabular}{@{}c@{}}\textbf{Vanilla}\\\vphantom{\textbf{Baseline}}\end{tabular} \\
                \midrule
                \multirow{3}{*}{Aspirin}
                                  & 5              & \textbf{50.20 $\pm$ 5.27$^*$}                                                    & 59.16 $\pm$ 5.47                                                   & \textbf{90.38 $\pm$ 9.38$^*$}  & 94.15 $\pm$ 10.58         \\
                                  & 10             & \textbf{31.84 $\pm$ 2.13$^*$}                                                    & 35.08 $\pm$ 4.22                                                   & \textbf{64.29 $\pm$ 3.98$^*$}  & 65.41 $\pm$ 4.46          \\
                                  & 20             & \textbf{23.06 $\pm$ 1.19}                                                        & 24.49 $\pm$ 1.32                                                   & \textbf{48.46 $\pm$ 0.92}      & 48.78 $\pm$ 0.61          \\
                \midrule
                \multirow{3}{*}{Azobenzene}
                                  & 5              & \textbf{26.85 $\pm$ 5.30$^*$}                                                    & 28.73 $\pm$ 7.80                                                   & \textbf{56.59 $\pm$ 3.38}      & 57.11 $\pm$ 3.11          \\
                                  & 10             & \textbf{14.74 $\pm$ 1.66}                                                        & 14.92 $\pm$ 1.85                                                   & 38.69 $\pm$ 2.84               & \textbf{38.60 $\pm$ 2.81} \\
                                  & 20             & \textbf{9.10 $\pm$ 0.27}                                                         & 9.38 $\pm$ 0.47                                                    & 25.74 $\pm$ 1.27               & \textbf{25.69 $\pm$ 1.50} \\
                \midrule
                \multirow{3}{*}{Malonaldehyde}
                                  & 5              & \textbf{34.27 $\pm$ 4.05$^*$}                                                    & 39.13 $\pm$ 4.02                                                   & \textbf{112.10 $\pm$ 4.53$^*$} & 113.96 $\pm$ 5.48         \\
                                  & 10             & \textbf{22.96 $\pm$ 1.75$^*$}                                                    & 25.19 $\pm$ 3.72                                                   & \textbf{78.64 $\pm$ 2.94$^*$}  & 79.34 $\pm$ 2.85          \\
                                  & 20             & \textbf{14.94 $\pm$ 1.59}                                                        & 15.36 $\pm$ 1.93                                                   & \textbf{57.02 $\pm$ 2.26}      & 58.40 $\pm$ 5.15          \\
                \midrule
                \multirow{3}{*}{Toluene}
                                  & 5              & 5.22 $\pm$ 1.26                                                                  & \textbf{5.10 $\pm$ 1.20}                                           & 15.45 $\pm$ 0.85               & \textbf{15.31 $\pm$ 0.72} \\
                                  & 10             & \textbf{3.12 $\pm$ 0.26}                                                         & 3.47 $\pm$ 1.02                                                    & 10.79 $\pm$ 0.56               & \textbf{10.65 $\pm$ 0.66} \\
                                  & 20             & 1.82 $\pm$ 0.08                                                                  & \textbf{1.80 $\pm$ 0.07}                                           & 7.41 $\pm$ 0.20                & \textbf{7.37 $\pm$ 0.22}  \\
                \bottomrule
            \end{tabular}%
        }
    \end{minipage}

\end{table}

\section{Limitations and Future Work}

Prototype-guidance has three main limitations. First, prototypes are constructed from energy representations only; incorporating force-aware prototypes could improve force MAE consistency. Second, our current evaluation is restricted to organic molecules; generalization to periodic or crystalline systems remains to be explored. These two are prime candidate for extending the framework in the Future. Third, the method requires access to source latent representations at fine-tuning time, which may be restrictive in settings where the pre-training dataset is unavailable.

\section{Conclusions}

We introduced a prototype-guided alignment framework for low-data fine-tuning of molecular machine-learning interatomic potentials (MLIPs). By constructing atom-type prototypes from the source training data and encouraging target representations to align with structurally analogous source environments, our method provides structured inductive bias without imposing a fixed functional form on the target potential energy surface.
Across all evaluated settings; SchNet and MACE on ST1 and ST2, and MACEOFF23 fine-tuned on seven compounds, prototype-guided latent alignment consistently matches or outperforms vanilla fine-tuning and rarely exhibits practically meaningful degradation. Energy MAE improvements are most pronounced in the lowest-data regime (5 training examples), where gains reach $\approx$14\% for MACE on Aspirin and $\approx$13\% for SchNet on Malonaldehyde, and are narrow but persist at larger system sizes. Force MAE improvements are present but smaller in magnitude and less uniform, suggesting that the prototype signal is most directly informative for energy learning, motivating future work in development of force prototypes, with current force benefits arising indirectly through shared representations. In the broader MACEOFF23 evaluation, where the strong pre-trained prior narrows the performance gap, prototype-guidance remains non-degrading and provides measurable energy improvements across several compounds, with no practical deterioration, confirming it as a generally superior fine-tuning strategy.

Overall, this work positions prototype guided latent alignment as a promising direction for enhancing data-efficient MLIP adaptation, potentially lowering the barrier to applying learned potentials in new molecular systems and supporting more efficient exploration of the chemical space.


\bibliographystyle{plainnat}
\bibliography{references.bib}


\appendix

\newpage
\section{Detailed Description of Prototype Alignment for MACE}

We provide a detailed overview of prototype alignment for a MACE model with two interaction blocks, illustrated in Figure~\ref{fig:data_graphs}.

Given an input molecular graph $\mathcal{G}^{(i)}$, atomic species are first converted into learnable embeddings and processed through successive interaction and product layers. From the model, we extract two types of latent representations for each atom $j$:

\begin{enumerate}
    \item \textbf{Interaction Latents:}
          Scalar features are extracted from the outputs of Interaction0 and Interaction1 and concatenated to form the interaction latent
          \[
              \mathbf{h}^{(i)}_j
              =
              \mathrm{Concat}
              \left(
              \mathbf{h}^{(i,0)}_j,
              \mathbf{h}^{(i,1)}_j
              \right).
          \]
          This representation captures multi-scale structural and geometric information about the atomic neighborhood.

    \item \textbf{Nodal Energy Contribution Latents:}
          Each readout layer predicts intermediate atomic energy contributions, which are summed to obtain the final atomic energy:
          \[
              E^{(i)}_j = \sum_k E^{(i,k)}_j.
          \]
          The stacked intermediate energy contributions form the nodal energy latent $\mathbf{e}^{(i)}_j$, encoding task-relevant energetic information.
\end{enumerate}

Concatenation of the features produced by both the interaction blocks ensures that all
possible information is passed to our method for construction and further alignment via these prototypes.


\section{Effect of $N_{\text{align}}$ \& $\boldsymbol{\lambda_{\text{neal}}}$}
\label{app:effect_of_n_align_and_lambda_neal}
Recall the $N_{\text{NEAL}}$ and $\lambda_{\text{NEAL}}$  determine the number of epochs and the strength of the alignment loss\eqref{eq:neal_loss}.
We initially experimented on MACE models with lower values of the NEAL weight , $\lambda_{\text{NEAL}} \in \{10^{-1}, 10^{-3}, 10^{-5}\}$, in combination with higher values of \texttt{NEAL\_end\_epoch} (e.g., 200). These settings led to marginal improvements for $10^{-3}$ and $10^{-5}$, while $10^{-1}$ degraded performance.

We then observed via t-SNE \cite{van2008visualizing} plots of interaction latents that most of their reorganization happens within the first few epochs of the model finetuning. We hypothesize that this behavior arises from the high degrees of freedom in the system, which causes the optimization trajectory to rapidly commit to the neighborhood of a particular local minimum during the initial training epochs. Subsequent epochs then primarily refine the parameters within this already selected basin of attraction, gradually converging toward the predetermined minimum.

Since the purpose of the prototype alignment framework is to guide the model to a better local optima, the above observation suggested that the loss be applied during the initial few epochs of the finetuning phase.

\begin{algorithm}[h!]
    \caption{Interaction Prototype Extraction using PCA+GMM}
    \label{alg:pca_gmm_prototypes}
    \begin{algorithmic}[1]
        \Require Foundation Model (FM), source-domain training subset $\mathcal{D}_s$, variance threshold $\tau_{\mathrm{var}}$, GMM component range $\mathcal{N}$
        \Ensure Interaction-space prototypes $\{(\boldsymbol{\mu}_I^{(k)}, \boldsymbol{\Sigma}_I^{(k)})\}$

        \State Extract interaction latent vectors $\{\mathbf{h}_j\}$ for each atom $j$ in $\mathcal{D}_s$ using the FM
        \State Fit PCA on $\{\mathbf{h}_j\}$ to retain at least $\tau_{\mathrm{var}}$ variance
        \State Project all $\mathbf{h}_j$ into reduced space: $\tilde{\mathbf{h}}_j \leftarrow \mathrm{PCA}(\mathbf{h}_j)$

        \For{all compound--element pairs $(c, z)$}
        \State Collect projected latents $\{\tilde{\mathbf{h}}^{(c,z)}_j\}$ for atoms of atomic number $z$ in compound $c$
        \For{all $n \in \mathcal{N}$}
        \State Fit a GMM with $n$ components on $\{\tilde{\mathbf{h}}^{(c,z)}_j\}$
        \State Compute BIC score
        \EndFor
        \State $n^\star \leftarrow \arg\min_{n \in \mathcal{N}} \mathrm{BIC}(n)$
        \State Extract $\{(\boldsymbol{\mu}_I^{(k)}, \boldsymbol{\Sigma}_I^{(k)})\}_{k=1}^{n^\star}$ from the best GMM
        \EndFor

        \State \Return All extracted interaction prototypes $\{(\boldsymbol{\mu}_I^{(k)}, \boldsymbol{\Sigma}_I^{(k)})\}$
    \end{algorithmic}
\end{algorithm}

\begin{algorithm}[h!]
    \caption{Energy Prototype Extraction from Interaction Clusters}
    \label{alg:energy_prototype_extraction}
    \begin{algorithmic}[1]
        \Require Foundation model (FM), source-domain training subset $\mathcal{D}_s$, interaction clusters $\{ \mathcal{C}_{c,z,j} \}$
        \Ensure Energy-space prototypes $\{ (\boldsymbol{\mu}_E^{(c,z,j)}, \boldsymbol{\Sigma}_E^{(c,z,j)}) \}$

        \State Extract nodal energy contribution latents $\{ \mathbf{e}_i \}$ for each atom $i \in \mathcal{D}_s$ using the FM

        \For{all compound--element pairs $(c, z)$}
        \For{all clusters $j$ with atoms in $\mathcal{C}_{c,z,j}$}
        \State Collect energy latents $\{ \mathbf{e}_i^{(c,z,j)} \}$ corresponding to atoms in cluster $\mathcal{C}_{c,z,j}$
        \State Fit a Gaussian distribution to $\{ \mathbf{e}_i^{(c,z,j)} \}$
        \State Extract mean vector $\boldsymbol{\mu}_E^{(c,z,j)}$ and covariance matrix $\boldsymbol{\Sigma}_E^{(c,z,j)}$
        \EndFor
        \EndFor

        \State \Return All extracted energy prototypes $\{ (\boldsymbol{\mu}_E^{(c,z,j)}, \boldsymbol{\Sigma}_E^{(c,z,j)}) \}$
    \end{algorithmic}
\end{algorithm}

\newpage

\section{Hyperparameter Details}
\label{hyperparameter-details}
We use the following hyperparameter configs for our experiments on SchNet.

\begin{table}[t]
    \centering
    \caption{Generic fine-tuning hyperparameters used for SchNet experiments on ST2.}
    \label{tab:schnet_finetuning_hparams}
    \begin{tabular}{ll}
        \toprule
        \textbf{Hyperparameter}        & \textbf{Value}                                     \\
        \midrule
        Model                          & SchNet                                             \\
        Optimizer                      & Adam                                               \\
        Learning rate                  & $\{10^{-3},\,10^{-4},\,5\times10^{-5},\,10^{-5}\}$ \\
        Weight decay                   & $5\times10^{-7}$                                   \\
        Batch size                     & [(5:2),(10:4),(20:8),(40:16)]                      \\
        Training epochs                & 1000                                               \\
        Early stopping patience        & 100                                                \\
        EMA decay                      & 0.99                                               \\
        Energy loss weight             & 8                                                  \\
        Force loss weight              & 1000                                               \\
        Training data shuffle          & True                                               \\
        Random seeds                   & $\{11,\,57,\,67\}$                                 \\
        Number of fine-tuning examples & $\{5\}$                                            \\
        NEAL end epoch                 & $\{10,\,20,\,30,\,40,\,50\}$                       \\
        NEAL alignment weight          & $\{1,\,10,\,100,\,1000\}$                          \\
        Hidden channels                & 256                                                \\
        Number of filters              & 256                                                \\
        Interaction blocks             & 3                                                  \\
        \bottomrule
    \end{tabular}
\end{table}
The vanilla finetuning methods use the same hyperparameters as the corresponding runs with NEAL, except NEAL-specific hyperparameters, NEAL end epoch, and NEAL alignment weight for both ST1 and ST2.

All experiments are averaged over 5 random seeds.

\begin{table}[htbp]
    \centering
    \caption{Hyperparameters used for training the Universal Model.}
    \label{tab:universal_model}
    \begin{tabular}{lc}
        \toprule
        \textbf{Hyperparameter}                   & \textbf{Value}     \\
        \midrule
        Number of interactions                    & 2                  \\
        Number of channels                        & 256                \\
        Maximum angular momentum ($\text{max}_L$) & 3                  \\
        Correlation                               & 3                  \\
        Maximum cutoff radius ($r_{\text{max}}$)  & 6.0                \\
        Batch size                                & 16                 \\
        Maximum number of epochs                  & 3500               \\
        Early stopping patience                   & 200                \\
        Learning rate (lr)                        & $1 \times 10^{-2}$ \\
        Weight decay                              & $5 \times 10^{-7}$ \\
        Force loss weight                         & 1000               \\
        Energy loss weight                        & 8                  \\
        SWA start epoch                           & 2000               \\
        SWA learning rate                         & $1 \times 10^{-3}$ \\
        SWA forces weight                         & 1000               \\
        SWA energy weight                         & 8                  \\
        EMA decay                                 & 0.99               \\
        \bottomrule
    \end{tabular}
\end{table}

\begin{table}[htbp]
    \centering
    \caption{Hyperparameters used for fine-tuning with IPR and AL trainable layers schemes across different finetuning techniques.}
    \label{tab:finetune_ipr_al}
    \begin{tabular}{lc}
        \toprule
        \textbf{Hyperparameter}           & \textbf{Value}             \\
        \midrule
        (Training budget, Batch size)     & [(5, 2), (10, 4), (20, 8)] \\
        Maximum number of epochs          & 1000                       \\
        Early stopping patience           & 80                         \\
        Learning rate (lr) Readout layers & $5 \times 10^{-3}$         \\
        Learning rate (lr) Other layers   & $1 \times 10^{-3}$         \\
        Weight decay                      & $5 \times 10^{-7}$         \\
        Force loss weight                 & 1000                       \\
        Energy loss weight                & 8                          \\
        SWA start epoch                   & 700                        \\
        SWA learning rate                 & $5 \times 10^{-4}$         \\
        SWA forces weight                 & 100                        \\
        SWA energy weight                 & 20                         \\
        EMA decay                         & 0.99                       \\
        \bottomrule
    \end{tabular}
\end{table}

\begin{table}[htbp]
    \centering
    \caption{Hyperparameters used for fine-tuning with RO trainable layers scheme across different finetuning techniques.}
    \label{tab:finetune_ro}
    \begin{tabular}{lc}
        \toprule
        \textbf{Hyperparameter}           & \textbf{Value}             \\
        \midrule
        (Training budget, Batch size)     & [(5, 2), (10, 4), (20, 8)] \\
        Maximum number of epochs          & 2000                       \\
        Early stopping patience           & 80                         \\
        Learning rate (lr) Readout layers & $1 \times 10^{-2}$         \\
        Weight decay                      & $5 \times 10^{-7}$         \\
        Force loss weight                 & 1000                       \\
        Energy loss weight                & 8                          \\
        SWA start epoch                   & 1500                       \\
        SWA learning rate                 & $5 \times 10^{-4}$         \\
        SWA forces weight                 & 100                        \\
        SWA energy weight                 & 20                         \\
        EMA decay                         & 0.99                       \\
        \bottomrule
    \end{tabular}
\end{table}
Fine Tuning of the universal model also uses similar hyperparameters to~\ref{tab:finetune_ipr_al},~\ref{tab:finetune_ro},~\ref{tab:universal_model} except parameters coming from the model such as number of channels is used.

\newpage

\FloatBarrier
\section{SMILES Strings for MACEOFF23 Compounds}
\label{appendix:smiles}

...

\begin{enumerate}
    \item \textbf{104069541}: \\
    \small\ttfamily
    \url{[N:1]\#[C:6][c:11]1[c:12]([N+:17](=[O:4])[O-:5])[c:14]([H:19])[c:16]2[c:15]([c:13]1[H:18])=[N:7][C:9](=[O:2])[C:10](=[O:3])[N:8]=2}
    \normalsize\normalfont

    \item \textbf{135030663}: \\
    \small\ttfamily
    \url{[O:1]=[C:5]1[N:3]=[N:4][C:6](=[O:2])[C:12]2([H:18])[C:10]([H:16])=[C:8]([H:14])[C:7]([H:13])=[C:9]([H:15])[C:11]12[H:17]}
    \normalsize\normalfont

    \item \textbf{135094581}: \\
    \small\ttfamily
    \url{[C:1]1([H:8])=[C:2]([H:9])[C:7]2([H:14])[C:4]([H:11])=[C:3]([H:10])[C:6]1([H:13])[N:5]2[H:12]}
    \normalsize\normalfont

    \item \textbf{135095347}: \\
    \small\ttfamily
    \url{[C:1](=[C:4]([C:7]([H:13])([H:14])[H:15])[C:8]([H:16])([H:17])[H:18])[C:2](=[C:5]([C:9]([H:19])([H:20])[H:21])[C:10]([H:22])([H:23])[H:24])[C:3]=[C:6]([C:11]([H:25])([H:26])[H:27])[C:12]([H:28])([H:29])[H:30]}
    \normalsize\normalfont

    \item \textbf{135161703}: \\
    \small\ttfamily
    \url{[C:1]1([H:15])=[C:2]([H:16])[C:5]2=[c:11]3[c:9]([H:21])[c:7]([H:19])[c:8]([H:20])[c:10]([H:22])[c:12]3=[C:6]3[C:13]([H:23])([H:24])[C:14]23[C:4]([H:18])=[C:3]1[H:17]}
    \normalsize\normalfont

    \item \textbf{135170767}: \\
    \small\ttfamily
    \url{[O:1]=[C:7]([O:6][C:10]([H:19])([H:20])[H:21])[C:16]12[C:12]3([H:23])[C:14]4([H:25])[C:11]5([H:22])[C:15]([H:26])([C:13]1([H:24])[C:17]53[N+:8](=[O:2])[O-:4])[C:18]42[N+:9](=[O:3])[O-:5]}
    \normalsize\normalfont

    \item \textbf{135246570}: \\
    \small\ttfamily
    \url{[O:1]1[C:4]2=[C:2]([H:15])[C:3]([H:16])=[C:5]([H:17])[C:14]([H:23])([H:24])[C:7]2=[c:13]2[c:11]([H:22])[c:9]([H:20])[c:8]([H:19])[c:10]([H:21])[c:12]2=[C:6]1[H:18]}
    \normalsize\normalfont
\end{enumerate}

\FloatBarrier
\newpage

\end{document}